# Band filling and cross quantum capacitance in ion gated semiconducting transition metal dichalcogenide monolayers

*Haijing Zhang[1, 2] \*, Christophe Berthod[1], Helmuth Berger[3], Thierry Giamarchi[1] and Alberto F. Morpurgo[1,2] \**

[1]*DQMP, University of Geneva, 24 Quai Ernest-Ansermet, CH-1211 Geneva, Switzerland*

[2]*GAP, University of Geneva, 24 Quai Ernest-Ansermet, CH-1211 Geneva, Switzerland*

[3]*Institut de Physique de la Matière Complexe, École Polytechnique Fédérale de Lausanne, CH-1015 Lausanne, Switzerland*

**Abstract:** Ionic liquid gated field-effect transistors (FETs) based on semiconducting transition metal dichalcogenides (TMDs) are used to study a rich variety of extremely interesting physical phenomena, but important aspects of how charge carriers are accumulated in these systems are not understood. We address these issues by means of a systematic experimental study of transport in monolayer $MoSe_2$ and $WSe_2$ as a function of magnetic field and gate voltage, exploring accumulated densities of carriers ranging from approximately $10^{14}$ cm$^{-2}$ holes in the valence band to $4 \cdot 10^{14}$ cm$^{-2}$ electrons in the conduction band. We identify the conditions when the chemical potential enters different valleys in the monolayer band structure (the K and Q valley in the conduction band and the two spin-split K-valleys in the valence band) and find that an independent electron picture describes the occupation of states well. Unexpectedly, however, the experiments show very large changes in the device capacitance when multiple valleys are occupied that are not at all compatible with the commonly expected quantum capacitance contribution of these systems, $C_Q = e^2 / \frac{d\mu}{dn}$. A theoretical analysis of all terms responsible for the total capacitance shows that –under general conditions– a term is present besides the usual quantum capacitance, which becomes important for very small distances between the capacitor plates. This term –that we call cross quantum capacitance– originates from screening of the electric field generated by charges on one plate from charges sitting on the other plate. The effect is negligible in normal capacitors, but large in ionic liquid FETs, because of the atomic proximity between the ions in the gate and the accumulated charges on the TMD, and it accounts for all our experimental observations. Our findings therefore reveal an important contribution to the capacitance of physical systems that had been virtually entirely neglected until now.

Ionic liquid gated field-effect transistors (FETs)[1-4] based on semiconducting transition metal dichalcogenides (TMDs) have been employed over the last several years to perform a plethora of interesting experiments, enabled by the possibility to accumulate unprecedented densities of charge carriers. The discovery that has attracted most attention is probably gate-induced superconductivity[5-11], which occurs upon accumulation of electron densities in excess of $n \sim 5 \cdot 10^{13}$ cm$^{-2}$. In the opposite regime of vanishingly low accumulated charge carrier density, the very large geometrical capacitance of ionic liquid gated devices has also led to interesting results, such as the possibility to determine the band gap of semiconducting TMDs quantitatively and precisely from simple transport measurements[12-16]. Other examples of interesting phenomena reported in TMD-based ionic liquid gated transistors include the observation of light emission from FETs operated in the ambipolar injection regime (i.e., with electron and holes injected at opposite contacts)[13, 17] or the enhancement of the electron-phonon interaction strength when multiple inequivalent valleys are filled by electrons[18, 19].

With all these fascinating phenomena having been observed, it is surprising that some seemingly simple aspects of ionic liquid gated TMD transistors continue to remain obscure. Indeed, in investigations of gate-induced superconductivity in different ionic liquid gated semiconducting TMDs, determining the position of the Fermi level in the band structure has proven difficult, and we are not aware of any experimental study in which the position of the chemical potential could be determined systematically as a function of accumulated carrier density (in fact, even in the highest quality TMD transistors based on conventional solid-state gates, interpreting low-temperature transport data to determine the position of the chemical potential is far from straightforward[20, 21]) . As a result, it is currently unclear which electronic states are responsible for the occurrence of superconductivity: whether it is the states in the valleys centered around the Q-point, around the K-point, or whether the occurrence of superconductivity requires multiple valleys to be occupied, as recent experimental and theoretical work suggests[18, 19, 22, 23]. Even the relation between the accumulated charge carrier density and applied gate voltage –i.e., the capacitance of ionic liquid gated TMD transistors– is not understood. It is routinely observed that the capacitance-per-unit-area of devices based on different TMDs –or even of layers of a same TMD, having different thicknesses– is significantly different despite the ionic liquid being the same[6, 12, 24]. As the very large geometrical capacitance of ionic liquids implies that quantum capacitance effects play a role, these differences have been attributed to differences in the density of states (DOS) of the TMD employed, without attracting particular attention. Whether the capacitance values measured in the experiments are consistent with the known properties of semiconducting TMDs remains, however, unclear.

Here we present a thorough analysis of charge accumulation in monolayer $MoSe_2$ and $WSe_2$ based on a quantitative analysis of longitudinal and transverse conductivity measurements as a function of gate voltage, for carrier density ranging from $n \sim 1.5 \cdot 10^{14}$ holes/cm$^2$ in the valence band to $n \sim 4 \cdot 10^{14}$ electrons/cm$^2$ in the conduction band. Upon accumulation of carriers in the conduction and valence band of $MoSe_2$, as well as in the conduction band of $WSe_2$, we observe a pronounced non-monotonic behavior in the measured longitudinal conductivity, accompanied by a sharp change in the Hall slope (i.e., the slope of the Hall resistance as a function of applied magnetic field $B$). Both phenomena are a manifestation of the occupation of a high energy valley: the longitudinal conductivity is affected through intervalley scattering, and the Hall slope because the occupation of a new valley changes the DOS and hence the total capacitance of the system. By using the values of effective masses known from either *ab-initio* calculations or angle-resolved photoemission spectroscopy (ARPES), we calculate the chemical potential $\mu(n)$ treating the accumulated carriers as independent particles. We find that –if the sharp change in Hall slope is taken as the indication that the chemical potential is entering a new valley– the energy offsets between the different valleys closely match the expected values. Unexpectedly, however, we also find that using the expression of $\mu(n)$ determined in this way, the measured capacitance $C$ of the device cannot be quantitatively accounted for by the series connection of the geometrical capacitance ($C_G$) and of the quantum capacitance ($C_Q = e^2 / \frac{d\mu}{dn}$), as commonly assumed. We argue that this is due to the existence of a cross quantum capacitance term that originates physically from mutual screening of charges on the two capacitor plates, since, indeed, charges in one layer (e.g., the charge carriers accumulated in the TMDs) screen the electric field generated by charges in the other layer (e.g., the charges at the surface of the ionic liquid). This term is predicted by theory (see Supplementary Information and Ref. [25]) and –although negligibly small in most systems studied in the past– it becomes important in our devices, because of the very close proximity between the charge accumulated in the TMD and the counter charge in the ionic liquid. Our results clarify aspects that are important to understand the filling of electronic states in semiconducting TMDs and reveal experimentally the microscopic nature of the capacitance of ionic liquid gated systems or –more generally– of systems with very small separation between the capacitor plates.

Our experiments rely on transport measurements performed on ionic liquid gated FETs realized on MoSe$_2$ and WSe$_2$ monolayers (see Fig. 1a). Monolayers were chosen because –in contrast to thicker multilayers– their band structure (Fig. 1b and 1c) is not strongly affected by the perpendicular electric field always present in a FET under charge accumulation conditions[26], which simplifies the interpretation of the experimental results. Measurements were performed at room temperature, so that the carrier density could be varied continuously by sweeping the gate voltage (this is not possible below $T \sim 280$ K, corresponding to the freezing temperature of P14-FAP, the ionic liquid used in our experiments). Fig. 1d shows an optical microscope image of the central part of one of our devices, consisting of a monolayer TMD contacted in a Hall bar configuration (see Supplementary Information for details of the fabrication process), prior to the deposition of the ionic liquid. A schematic representation of a complete device is shown in Fig. 1e, which includes also the gate electrode, the reference electrode, and the electrical bias scheme used to measure the longitudinal and transverse conductivity/resistivity. The reference electrode is necessary to perform a precise quantitative analysis, as it allows the potential difference $V_{\text{ref}}$ between the ionic liquid and the FET channel (i.e., the voltage across the ionic liquid gate capacitor) to be measured directly. This removes the effect of a possible spurious voltage drop at the interface between the gate electrode and the ionic liquid. Thus, in presenting the experimental results, data are often plotted as a function of $V_{\text{ref}}$ and not of $V_{\text{G}}$.

The linear conductivity of a MoSe$_2$ and WSe$_2$ monolayer liquid FETs measured upon sweeping the gate voltage exhibits clear ambipolar transport, as shown in Fig. 1f and 1g. Indeed, a large conductivity is observed if a sufficiently negative/positive gate voltage $V_{\text{G}}$ (see top axis) is applied to shift the chemical potential $\mu$ into the valence/conduction band of the TMD monolayers. In the $V_{\text{G}}$ interval where the conductivity vanishes, the chemical potential $\mu$ is located inside the band-gap, and the measurements allow the size of the gap to be determined.

In simple terms, a change in gate voltage (or more precisely in reference potential, $\Delta V_{\text{ref}}$) induces a change in chemical potential $\Delta\mu$ and in electrostatic potential $\Delta\varphi$, related as $e\Delta V_{\text{ref}} = \Delta\mu + e\,\Delta\varphi$. For a parallel plate capacitor, $\Delta\varphi$ can be expressed in terms of the accumulated charge density $\Delta n$ as $\Delta\varphi = e\,\Delta n/C_{\text{G}}$. When $\Delta\mu$ is shifted through the band gap, the change in charge density is very small, because there are nominally no states within the gap to accumulate charge. Together with the fact that the geometrical capacitance $C_{\text{G}}$ of ionic liquid gated devices is very large, this allows us to neglect $\Delta\varphi$ and write $e\Delta V_{\text{ref}} = \Delta\mu$. Since the threshold voltages for electron and hole accumulation ($V_{\text{th}}^{e}$ and $V_{\text{th}}^{h}$) correspond to having $\mu$ located respectively at the conduction and valence band edges, their difference directly measures the band-gap, i.e. $\Delta = e(V_{\text{th}}^{e} - V_{\text{th}}^{h})$. From measurements on devices based on MoSe$_2$ and WSe$_2$ monolayers we find $\Delta_{\text{MoSe}_2} = 1.98$ eV and $\Delta_{\text{WSe}_2} = 1.87$ eV. These values can be used to obtain the exciton binding energy by subtracting the known exciton recombination energy[27], from which we obtain 415 meV and 240 meV for MoSe$_2$ and WSe$_2$, respectively. All these quantities (band-gaps and exciton binding energies) are in excellent agreement with values reported in the literature[28-30], showing that ionic liquid gated devices are well suited to perform quantitative investigations of MoSe$_2$ and WSe$_2$ monolayers.

We now proceed to consider the regime of interest in this work, with the chemical potential $\mu$ located inside either the conduction or the valence band. In this case the electrostatic potential cannot be neglected, which is why the determination of the chemical potential as a function of $V_{\text{ref}}$ becomes more complex. The data show that the evolution of the conductivity upon adding electrons ($V_{\text{G}} > 0$) is non-monotonic for both MoSe$_2$ and WSe$_2$; if holes are accumulated ($V_G < 0$), a non-monotonic behavior is also observed, but only in the MoSe$_2$ monolayer. In all cases the measurements are entirely reproducible upon sweeping $V_{\text{G}}$ multiple times, there is virtually no hysteresis, and the leakage current is vanishingly small: indeed all basic aspects of our observations have been reproduced in multiple devices, and the behavior shown in Fig. 1f and 1g is representative of ionic liquid gated MoSe$_2$ and WSe$_2$ monolayers (for measurements illustrating the level of reproducibility see the Supplementary Information).

Based on similar observations made earlier on ion-gated bi and trilayer graphene[31], as well as on *ab-initio* calculations of the transport properties of monolayer TMDs[32], we attribute the conductivity non-monotonicity to inter-valley scattering processes that become possible as the chemical potential is shifted into a higher energy valley. To search for additional experimental evidence indicating the occupation of a high energy valley, we have measured the Hall effect and analyzed the evolution of the carrier density extracted from transverse resistance (i.e., as $n = \left(e \frac{dR_{xy}}{dB}\right)^{-1}$) as a function of $V_{\text{ref}}$. Two types of measurements were done. First, we fixed the applied gate voltage, measured the transverse resistance as a function of magnetic field, antisymmetrized the results to eliminate any possible offset, and extracted the Hall slope. Results of selected measurements performed on a MoSe$_2$ monolayer are shown in the left and right panels of Fig. 2a, respectively in the regime of hole or electron accumulation. In a second type of measurement we fixed the magnetic field $B$ first at $+5$ T and then at $-5$ T, measured in each case $R_{xy}$ as a function of $V_{\text{ref}}$, and extracted the Hall slope as $\frac{dR_{xy}}{dB} = \frac{R_{xy}(5\text{ T}) - R_{xy}(-5\text{ T})}{10\text{ T}}$. Figure 2b shows that the electron densities in the conduction band of MoSe$_2$ obtained following these two procedures coincide for all values of $V_{\text{ref}} - V_{\text{th}}^e$, illustrating the excellent stability and reproducibility of our devices (see Supplementary Information, Fig. S2, for data measured on WSe$_2$, analogous to those shown in Fig. 2a and 2b for MoSe$_2$).

For MoSe$_2$ and WSe$_2$ monolayers, the carrier density extracted from Hall effect measurements is plotted in Fig. 2c and 2d throughout the entire $V_{\text{ref}}$ range explored. In the valence and conduction band of MoSe$_2$, as well as in the conduction band of WSe$_2$, the relation between $n$ and $V_{\text{ref}}$ is non-linear: $n$ increases linearly with $V_{\text{ref}}$ up to a density $n^*$ (whose value depends on material and band considered: $n^* \cong 3.0 \cdot 10^{13}$ and $1.0 \cdot 10^{14}$ electrons/cm$^2$ for the conduction band of WSe$_2$ and MoSe$_2$, and $n^* \cong 6.3 \cdot 10^{13}$ holes/cm$^2$ for the valence band of MoSe$_2$), past which $n$ continues to increase linearly with a slope that is significantly steeper, directly indicating a change in the device capacitance for $n > n^*$. Only in the valence band of WSe$_2$ – i.e., the sole case in which the non-monotonicity of the conductivity was absent– a slope change in the $n$-vs-$V_{\text{ref}}$ relation is not present. This observation confirms that the non-monotonic behavior of the conductivity and the change in device capacitance have the same microscopic origin.

To ensure that the peculiar gate voltage dependence of the conductivity and of the carrier density observed in our experiments is not an artefact due to the ionic liquid that we have selected, we have performed additional measurements using a different ionic liquid. The inset of Fig. 2d shows the dependence of the electron density (orange dots) and of the conductivity (red line) measured on a monolayer WSe$_2$ FET realized using DEME-TFSI. It is clear from the data that the observed behavior is identical to that of devices based on P14-FAP: the longitudinal conductivity exhibits a pronounced non-monotonicity and the slope of the electron density as a function of $V_{\text{ref}}$ changes abruptly for $n = n^*$, with the value of $n^*$ that coincides in DEME-TFSI and P14-FAP-based devices within the experimental accuracy ($n^* \cong 3.0 \cdot 10^{13}$ electrons/cm$^2$ for P14-FAP and $3.2 \cdot 10^{13}$ electrons/cm$^2$ for DEME-TFSI). We conclude that the phenomenology that we observe is very robust experimentally: it is a manifestation of the electronic properties of the semiconducting monolayer TMD studied, and not of the ionic liquid employed.

At a qualitative level, the observed evolution of the conductivity and capacitance of MoSe$_2$ and WSe$_2$ monolayers appears to be internally consistent. The filling of a high-energy valley can account for a non-monotonicity in the conductivity (because of inter-valley scattering), and the DOS associated to the same valley for the increase in the device capacitance. Indeed, the total capacitance $C$ of field-effect devices is commonly viewed as due to the series connection of the geometrical capacitance and the quantum capacitance, $1/C = 1/C_{\text{G}} + 1/C_{\text{Q}}$: an increase in DOS directly causes an increase of $C$. As the change in capacitance measured experimentally is unexpectedly large (in the conduction band $C$ increases by approximately a factor of 3 as $n$ is increased past $n^*$), the question is whether such an interpretation does work at a quantitative level. As we discuss here below, the existing knowledge of the electronic structure of TMD monolayers coming from *ab-initio* calculations and ARPES experiments, together with the possibility to tune the carrier density to occupy different valleys, allow us to gain new, unexpected insights about the quantum capacitance of ionic liquid gated devices.

We start by analyzing whether filling of the conduction and valence bands as described in terms of independent electrons or holes is consistent with the experimental observations. For monolayer MoSe$_2$ and WSe$_2$ the structure of the conduction and valence band is known (see Fig. 3a). Upon adding electrons, the K-valley is filled first, and past a threshold density –that we identify with $n^*$ shown in Fig. 2c and 2d– also the Q-valley starts to be filled. To calculate $\mu(n)$, we need to know the DOS associated to both these valleys, determined by their degeneracy and effective mass. In the conduction band, states can be taken to be spin degenerate, because the magnitude of the spin-splitting due to spin-orbit interaction is comparable or smaller than the thermal energy (i.e., of $k_BT$ at room temperature). An additional orbital degeneracy is present, that is 2 for the K-valley and 6 for the Q-valley. We take the effective masses from *ab-initio* calculations, believed to be rather accurate (in the K-valley $m^* = 0.58\ m_0$ for MoSe$_2$ and $m^* = 0.33\ m_0$ for WSe$_2$; in the Q-valley $m^* = 0.78\ m_0$ for MoSe$_2$ and $m^* = 0.43\ m_0$ for WSe$_2$)[33], using which we readily calculate the chemical potential $\mu(n)$. The chemical potential obtained in this way for the electron density values shown in Fig. 2c and 2d is represented by the green circles in Fig. 3b and 3c. Because of the constant DOS, $\mu(n)$ increases linearly with density up to $n^*$. For $n > n^*$, the increase is still linear but with a smaller slope, due to the additional contribution to the DOS from the Q-valley; the value of $\mu$ for $n = n^*$, i.e. $\mu(n^*)$, corresponds to the distance in energy $E_{K-Q}$ between the conduction band edges at the Q and the K point. Using this criterion, we find that $E_{K-Q} = 205$ meV and 108 meV for MoSe$_2$ and WSe$_2$, in agreement with literature values. In particular, for MoSe$_2$ tunneling spectroscopy measurements indicate that $E_{K-Q} = 190$ meV [29], only 5% off from the value obtained through our analysis. For WSe$_2$ no direct measurements are currently available, but *ab-initio* calculations reported in the literature give values for $E_{K-Q}$ between 30 and 120 meV [26, 33, 34], consistent with the value extracted from our experimental analysis.

The same comparison can be performed for holes accumulated in the valence band, with the distinctive advantage that all key parameters (i.e., effective masses and energy distances between valleys) are known from ARPES measurements. In particular, holes are initially accumulated at the K-valley, whose states are not spin-degenerate because of the very large spin-splitting. If a sufficiently high hole density can be accumulated, the second valley to be filled depends on the material: in MoSe$_2$ it is the spin-split valley at K, whereas in WSe$_2$ –due to the very large magnitude of the spin-splitting in the valence band of this material– it is the valley centered at Γ. The measured hole mass at the K-valley is $m^* = 0.67\ m_0$ for MoSe$_2$ [35] and $m^* = 0.45\ m_0$ for WSe$_2$ [36, 37] (for both high spin and low spin), using which the chemical potential $\mu(n)$ can then be calculated as a function of hole density (see yellow circles in Fig. 3b and 3c). We find that in MoSe$_2$, the next valley starts being filled when $\mu(n) = 224$ meV, in good quantitative agreement with the known value of spin-orbit splitting at the K-point (240 meV) [37]. In WSe$_2$, no higher energy valley is filled in the experimentally accessible range of hole density. This is indeed consistent with the known properties of WSe$_2$ monolayers, since the maximum value of $\mu(n)$ reached in our experiments is smaller than 400 meV, whereas filling a second valley requires reaching values in excess of 500 meV [37]. We conclude that for both electrons and holes, under the conditions of our experiments, filling of states is satisfactorily described in terms of independent particles.

An important consequence of the analysis presented above is that the expression for the chemical potential $\mu = \mu(n)$ can be directly used to calculate the quantum capacitance, $C_Q = e^2/\frac{d\mu}{dn}$ [38-42]. We obtain $C_Q = e^2 \nu(\mu)$, with $\nu(\mu)$ the DOS at the chemical potential, as it should be expected, since our analysis shows that treating charge carriers as independent describes well band filling. The quantum capacitance, however, cannot be directly compared to the experimental data, because what is measured is the total capacitance $C$ given by $1/C = 1/C_G + 1/C_Q$, and the value of the geometrical capacitance is unknown. The issue can be readily solved because –if we fix the polarity of $V_G$, i.e., the band to be considered– the geometrical capacitance is expected to be the same for the two valleys and for the two materials. We can then determine the value of $C_G$ by imposing that in one of the materials at low density –i.e., when only the K-valley is occupied– the total capacitance $C = \frac{C_G C_Q}{C_G + C_Q}$ matches the measured value, and then check if with the same value of $C_G$ the capacitances measured at higher density and/or in the other material are correctly reproduced ($C_G$ can differ upon inverting the sign of $V_G$, because for opposite polarities the accumulated ions are different molecules, with different size).

We have carefully performed this type of comparison for all valleys of MoSe$_2$ and WSe$_2$, both in the conduction and valence band, and found that deviations –a factor of 2-to-3 as compared to the measured values– are systematically present. Therefore, the conventional expression for the quantum capacitance $C_Q = e^2/\frac{d\mu}{dn}$ does not reproduce the experiments quantitatively. This same analysis shows that if we do not take $C_Q$ to be equal to $e^2 \nu(\mu)$, but just proportional to it –i.e., we assume $C_Q^{\text{exp}} = \alpha e^2 \nu(\mu)$ with the same value of the proportionality constant $\alpha$ for a same polarity, irrespective of material and valley– the data can be reproduced quantitatively (see Fig. 4a and 4b). The geometrical capacitances that enable the data to be reproduced are 57 μF/cm² and 41 μF/cm² in the conduction and valence band respectively, very reasonable values for ionic liquids[43]. Fig. 4c shows that, with these $C_G$ values, the values of the experimentally determined quantum capacitance $C_Q^{exp}$ is proportional to the DOS with the same proportionality constant equal to $\alpha = 0.29$ in the conduction band and $\alpha = 0.47$ in the valence band, irrespective of valley and material (to be specific, the values of $C_G$ and of $\alpha$ are obtained by least-square fitting of the measured capacitance expressed as a function of the DOS as $\frac{1}{C} = \frac{1}{C_G} + \frac{1}{\alpha e^2 \nu(\mu)}$).

Finding that the experimentally determined value of quantum capacitance is $C_Q^{exp} = \alpha e^2 \nu(\mu)$ (with $\alpha < 1$) rather than $C_Q = e^2 / \frac{d\mu}{dn}$ implies that an additional, sizable contribution to the total capacitance proportional to $e^2 \nu(\mu)$ is present, which adds in series to the commonly considered geometrical and quantum terms. It may be argued that this additional contribution comes from interactions between carriers in the TMD, since correlation effects can crucially modify their total energy as compared to simply their kinetic energy. However, this possibility does not appear to be compatible with our observation that band filling is well described within an independent electron scenario. Indeed, as we discussed above, the experiments show that the dependence of $\mu(n)$ obtained from a description based on independent particles accounts systematically for the values of $n$ at which the occupation of higher energy valleys starts. Any significant modification of the quantum capacitance $C_Q = e^2 / \frac{d\mu}{dn}$ due to correlations would necessarily imply a large modification of $\frac{d\mu}{dn}$, and hence of $\mu(n)$, violating such an excellent agreement.

A realistic scenario compatible with all experimental observations is revealed by a theoretical analysis of other ways in which interactions determine the capacitance. All the key elements of this theoretical description are presented in the Supplementary Information and more details will be discussed in a forthcoming publication[25]. Under general conditions, theory shows that a capacitance term originating from interaction-induced cross-correlations between carriers accumulated on the opposite plates of a capacitor (as opposed to interactions between charges on a same plate mentioned here above) is always present. This term –that we refer to as *cross quantum capacitance* and that can be understood as a mutual screening effect– leads to a contribution to the total energy that cannot be written as due exclusively to carriers that are either on one or the other capacitor plate. Interaction-induced correlations between carriers located in different nearby conductors are known to occur[44] and are key to understand phenomena such as Coulomb drag[45-48] or plasmon dispersion in semiconducting heterostructures[49, 50]. Their influence on the capacitance has received little attention, because these cross correlations become strong only when the distance between the capacitor plates is comparable or less than the distance between charges inside the plates. The phenomenon is therefore not relevant in determining the capacitance of most devices, but it is for ours, due to the very close proximity of the ionic liquid to the TMD. Indeed, it appears physically realistic that the carriers accumulated on the TMD monolayer provide a determinant contribution to the screening of interactions between the excess charge in the ionic liquid, which is in direct contact with the monolayer itself.

To explain the origin and nature of the cross quantum capacitance we provide simple physical arguments (see Supplementary Information for a discussion of the underlying description based on linear-response theory; technical details –too lengthy to be discussed here– will be presented elsewhere[25]). To this end, Fig. 5 compares different situations, in which the total capacitance is dominated by the geometrical contribution (Fig. 5a), exhibits a sizable quantum capacitance term due to the kinetic energy contribution (Fig. 5b), or includes also a cross quantum capacitance contribution (Fig. 5c). In the first case, valid for conventional capacitors with a small geometrical capacitance (e.g., due to a sufficiently large distance between the plates), the gate bias $V_{\text{ref}}$ is entirely converted into an electrostatic potential difference $\Delta\varphi$ between the charges located on the opposite capacitor plates (i.e., $\Delta\varphi = V_{\text{ref}}$)[51]. It then follows directly from Gauss law ($\text{div}(\epsilon\vec{E}) = \rho$) that $\frac{V_{\text{ref}}}{d} = \frac{ne}{\epsilon}$ (where $ne$ is the density of charge accumulated on the plates), and consequently $C_G = \frac{\epsilon}{d}$ (i.e., the usual expression for the geometrical capacitance). In Fig. 5b, the positive charges are still classical, but the negative charges are subject to Pauli exclusion principle. As they pile up on the right capacitor plate, the chemical potential raises, so that the gate bias is shared between the electrostatic potential drop $\Delta\varphi$ and the shift in chemical potential $\Delta\mu$. In this case Gauss law reads $\frac{\Delta\varphi}{d} = \frac{V_{\text{ref}} - \Delta\mu/e}{d} = \frac{ne}{\epsilon}$ that –using the relation $n = \nu\Delta\mu$ for non-interacting particles in two dimensions– yields for the total capacitance $C$ the expression $1/C = 1/C_G + 1/C_Q$, with $C_Q = e^2\nu(\mu)$.

We can understand the origin of the cross quantum capacitance reasoning along the same lines. The cross quantum capacitance becomes important when the separation between the capacitor plates $d$ is sufficiently small, so that mutual screening of charges on the two separate capacitor plates cannot be neglected. A self-consistent theory (see Supplementary Information) predicts the existence of a wealth of microscopically distinct cross-screening processes, whose net effect is to reduce the amount of charge that is accumulated onto the capacitor plates by the applied gate bias, as schematically illustrated in Fig. 5c. The phenomenon can be described in terms of a screening potential $V_s$, proportional at leading order to the total accumulated charge, i.e. $V_s = \frac{ne}{C_s}$, with the constant $C_s$ that can –at least in principle– be calculated within the linear response theory. A finite $V_s$ causes the electrostatic potential difference between the two capacitor plates to be reduced as $\Delta\varphi = V_{\text{ref}} - V_s - \Delta\mu/e$, and accordingly Gauss law reads $\frac{V_{\text{ref}} - V_s - \Delta\mu/e}{d} = \frac{ne}{\epsilon}$. It directly follows that the constant $C_s$ represents an additional capacitance that adds in series to the geometrical capacitance and to the quantum capacitance.

These results can be obtained through a rigorous theoretical analysis of the charge susceptibility (i.e., the relative change of charge density in response to a change in electric potential) in a system formed by two spatially separated conductors –the capacitor plates– with Coulomb interaction acting between charges sitting on the same conductor and on different conductors (see Supplementary Information). Linear-response theory expresses the different contributions to the total capacitance in terms of the intra- and inter-plate contributions to the charge susceptibility. Within this scheme, approximate calculations relying on a perturbative analysis of the inter-plate susceptibility show that the cross quantum capacitance $C_s$ is proportional to the DOS. They also show that if the charges on one of the capacitor plates are frozen –i.e., they are not able to screen other charges– and charges on the other plate can be treated as independent particles, the cross quantum capacitance reduces the total quantum capacitance by up to a factor of 4 (depending on the distance between the plates) relative to the value $e^2\nu(\mu)$. These theoretical considerations are therefore fully consistent with the experimental data, namely with the fact that in ionic liquid gated monolayer TMD FETs an additional contribution proportional to $e^2\nu(\mu)$ is present that reduces by a factor of 2-to-3 the conventional quantum capacitance term. In other words, the quantity $C_Q^{\text{exp}}$ that we extract from our measurements does not correspond to the common quantum capacitance $C_Q$, but to the series connection of the quantum capacitance and of the cross quantum capacitance (i.e., $1/C_Q^{exp} = 1/C_Q + 1/C_s$).

The experimental results and theoretical considerations taken together provide a consistent picture of band filling in ionic liquid gated transistors based on monolayer semiconducting TMDs. Experimentally, it is the possibility to fill multiple bands that enables a precise analysis of both band filling and of the microscopic nature of the capacitance. Indeed, the change in capacitance as a function of gate bias marks the density at which the second valley starts to be filled, and is key to conclude that an independent particle description of band filling agrees with the data. Filling multiple valleys is also essential to analyze quantitatively the total capacitance of our devices. These two points have enabled the most relevant finding of our work –that has come entirely unanticipated– namely the identification of a sizable cross quantum capacitance contribution. As the notion of cross quantum capacitance had remained virtually unappreciated until now, these results disclose a number of possibilities for future work. For instance, in capacitors formed by graphene electrodes separated by extremely thin boron nitride exfoliated crystals, cross quantum capacitance effects may also be expected. Indeed, recent measurements on this type of devices show deviations of the total capacitance from theoretical predictions considering only conventional quantum capacitance effects[42]. The deviations are significant only at sufficiently low density, compatibly with the notion that the cross quantum capacitance contribution becomes important when the average distance between carriers is smaller than the electrode separation. An analysis of this regime may therefore allow details of the cross quantum capacitance to be better understood microscopically, because the electronic properties of graphene electrodes are easier to describe than those of ionic liquids. For ionic liquid gated FETs the results obtained here are of great interest as they may lead to device improvements (e.g., the accumulation of even larger carrier density as compared to what is already possible now) and better device control. For instance, given the very large reduction of the total capacitance due to cross quantum capacitance effects, it may be even expected that introducing a very thin (e.g., mono, bi, trilayer) hBN crystal between the TMD and the ionic liquid[52] leads to a capacitance increase. This could happen because increasing the capacitor thickness can suppress the cross capacitance term more rapidly than the other contributions to the total capacitance. The observation of such a counterintuitive phenomenon would provide direct experimental evidence for the presence of a capacitance term beyond the geometrical and the conventional quantum capacitance contributions. More ideas and implications of cross capacitance can be easily imagined, for instance for the accumulation of ions at the surface of metallic electrodes in batteries[53]. It is clear that more experimental and theoretical work to explore all these phenomena is needed in order to reach a complete microscopic understanding of cross quantum capacitance.

**Corresponding Authors**

*Email: haijing.zhang@unige.ch

*Email: alberto.morpurgo@unige.ch

**Acknowledgements**

We gratefully acknowledge A. Ferreira for continued technical support of the experiments. Financial support from the Swiss National Science Foundation and the EU Graphene Flagship project is also gratefully acknowledged.

present, the applied gate bias partly consists of an electrostatic potential and partly of a shift in the chemical potential of the electrons in the capacitor plates.

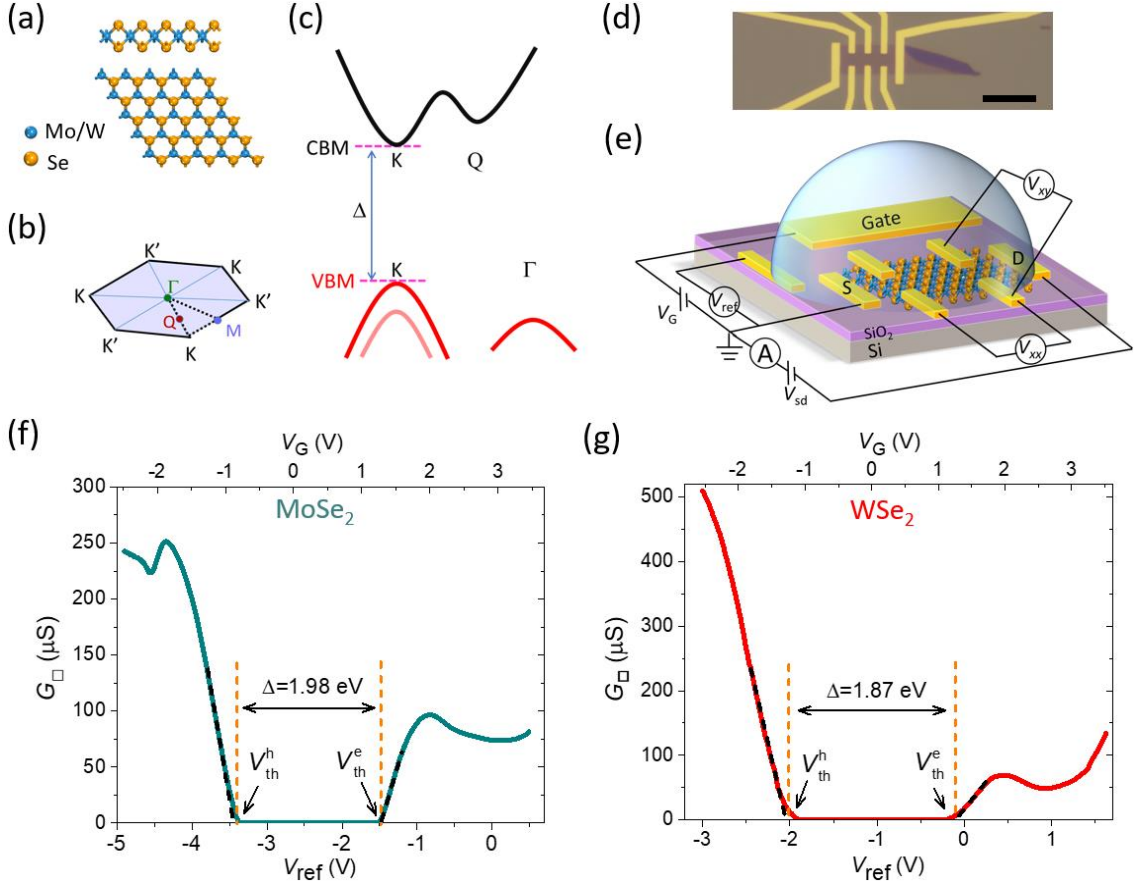

**Figure 1.** (a) Schematic structure of monolayer $MoSe_2$ and $WSe_2$. The corresponding Brillouin zone is represented in (b), showing the points in reciprocal space around which valleys relevant for our work are centered. (c) Schematic illustration of the conduction and valence band edges of monolayer $MoSe_2$ and $WSe_2$. The conduction band edge consists of a single line because at the temperature of our experiments spin-splitting is smaller than the thermal energy and can be neglected. This is not the case for the valence band edge, since in the valence band spin-orbit interaction is much larger than $k_BT$ (the light-red line in the valence band corresponds to the spin-split K-valley; in $WSe_2$ the high-spin-split K valley is higher in binding energy than the $\Gamma$-valley; see main text). (d) Optical microscope image of a monolayer $MoSe_2$ contacted in a Hall bar configuration (the scale bar is 10 μm). (e) Full schematics of an ionic liquid gated FET, showing also gate and reference electrodes, as well as the electrical circuit used to bias and measure the device. (f) and (g) show the dependence of the square conductance (or, equivalently, conductivity) of $MoSe_2$ and $WSe_2$ as a function of either reference potential $V_{ref}$ (lower axis) or gate voltage $V_G$ (upper axis).

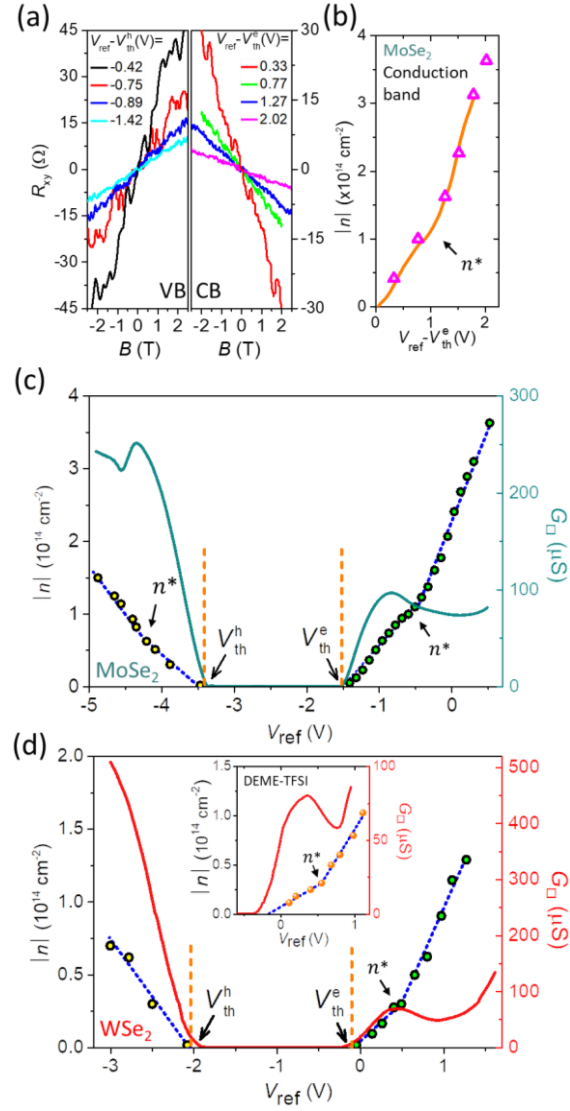

**Figure 2.** (a) $B$ dependence of the Hall resistance of MoSe$_2$ monolayer measured for different applied gate voltages (see values of $V_{ref}$ in the legend), to shift the chemical potential in the valence band (left panel) and in the conduction band (right panel). (b) Carrier density extracted from the slope of the Hall resistance measured while sweeping $B$ at fixed $V_{ref}$ (empty triangles), or while sweeping $V_{ref}$ at fixed $B$ (orange continuous line; see main text). The excellent agreement demonstrates the high stability of our devices. The evolution of carrier density (left axis) and longitudinal conductivity (right axis) as a function of $V_{ref}$ is compared in (c) and (d), respectively for monolayer MoSe$_2$ and WSe$_2$. Irrespective of material and band, a non-monotonic dependence of the conductivity occurs if and only if a non-linearity in the dependence of $n$ on $V_{ref}$ (setting in at $n^*$) is observed, showing that the two phenomena have the same origin (green and yellow circles represent respectively electron and hole density). Inset in (d): data taken on a WSe$_2$ monolayer device using DEME-TFSI as ionic liquid (instead of P14-FAP) show that the observed behavior does not depend on the specific liquid used.

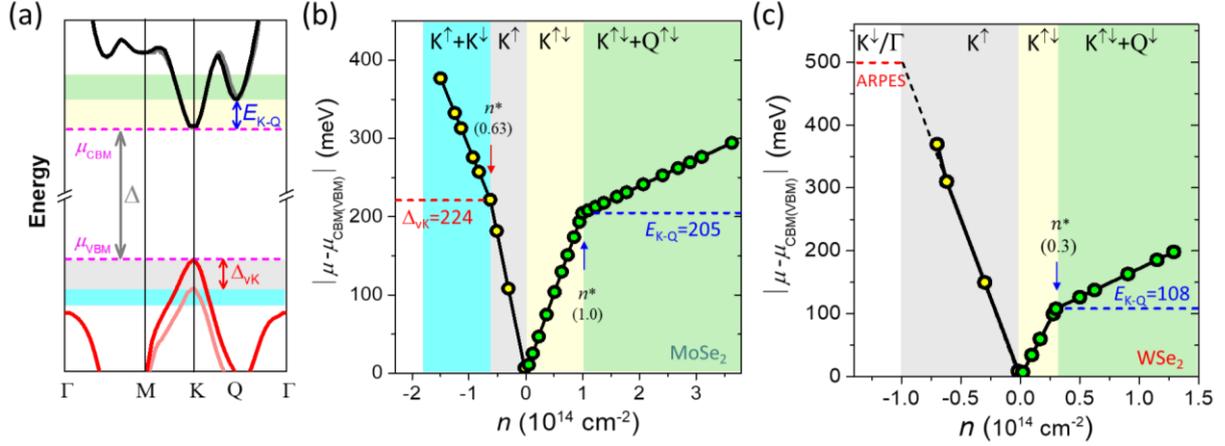

**Figure 3.** (a) Schematic representation of the conduction and valence bands of semiconducting TMDs showing the K- and Q-valleys in the conduction band, as well as the (spin-split) K-valley and the Γ-valley in the valence band ($E_{K-Q}$ indicates the energy distance between the conduction band edges at the K- and Q-valleys; $\Delta_{vK}$ is the magnitude of the spin-splitting of the K-valley in the valence band). Shadings of different colors represent the intervals of chemical potential corresponding to having different configurations of filled valleys: in the conduction band, light-yellow corresponds to having only the K-valley filled, and green to having filled both the K- and the Q-valleys; in the valence band, for grey only the low-spin valley at K is filled and for light-blue both the low- and the high-spin valleys are filled. The same shadings are used in panels (b) and (c) to identify the corresponding intervals of carrier density. Panels (b) and (c) represent the position of the chemical potential relative to the conduction (green circles) or valence (yellow circles) band edge, as a function of carrier density (negative values correspond to holes; the value of $\mu$ is extracted as explained in the main text). In (c) the dashed line marked "ARPES" indicates the value of the chemical potential at which a new valley (centered at the Γ-point, as established by ARPES measurements) would be populated. The required hole density is too high to be reached in the experiments, which is why the occupation of a second valley in WSe$_2$ is not observed experimentally upon hole accumulation.

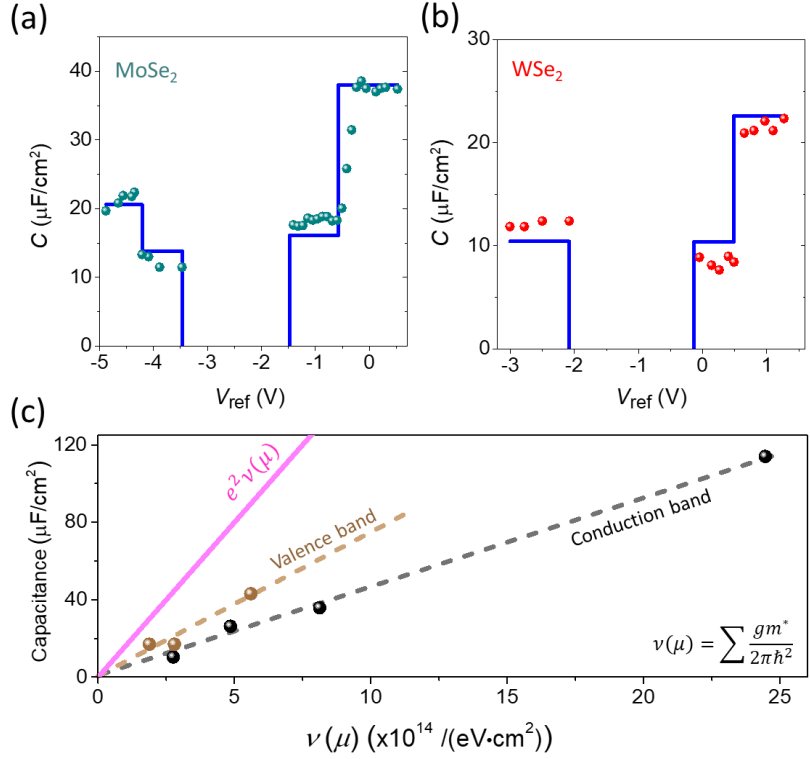

**Figure 4.** The symbols in panels (a) and (b) represent the experimentally determined capacitance values of MoSe$_2$ and WSe$_2$ devices, as extracted from Hall resistance measurements, at different reference voltages $V_{\text{ref}}$ ( $C = \frac{|ne|}{V_{\text{ref}}}$ ). The continuous lines are calculated using the formula $\frac{1}{C} = \frac{1}{C_G} + \frac{1}{C_Q^{\text{exp}}}$, with $C_Q^{\text{exp}} = \alpha\, e^2 \nu(\mu)$ ($\nu(\mu)$ is the value of the DOS at the chemical potential $\mu$, calculated using the effective masses and degeneracies of the occupied valleys). The values of geometrical capacitance $C_G$ and $\alpha$ for a given band are the same irrespective of valley and material, and are chosen to optimize the agreement with the measured data. The result of this analysis is illustrated in panel (c) by plotting the capacitance contributions $C_Q^{\text{exp}}$ as a function of the density of states $\nu(\mu)$: full dots are experimental points and dashed lines indicate their linear dependence on DOS (brown and black symbols/lines correspond to the valence and conduction band, respectively). The pink line represents the common expression for the quantum capacitance of independent electrons $C_Q = e^2 \nu(\mu)$. The clear and pronounced suppression of $C_Q^{\text{exp}}$ as compared to $C_Q = e^2 \nu(\mu)$ (by a factor $\alpha = 0.47$ in the valence band and $\alpha = 0.29$ in the conduction band) is a manifestation of the cross quantum capacitance term discussed in the main text.

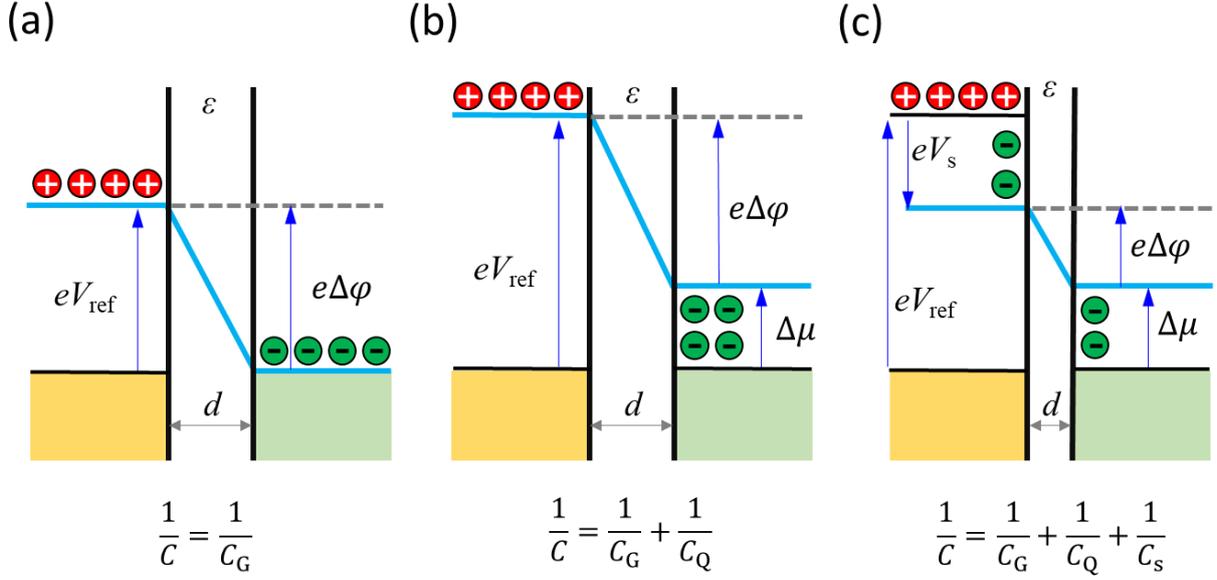

**Figure 5.** Classical and quantum capacitors. (a) In a capacitor made of classical charges, with a sufficiently large separation between the plates, the applied bias is entirely converted into an electrostatic energy difference between the positive and negative charges and the capacitance is the geometrical capacitance. (b) If the negative charges obey Fermi statistics, part of the applied bias is consumed for a shift of chemical potential: a larger bias is needed for accumulating the same amount of charges as in (a), hence the capacitance is reduced. This is the origin of the quantum capacitance term connected in series with the geometrical one. (c) For ultrathin capacitors, the plates screen each other's charges, building a screening potential $V_s$ that opposes to the applied bias, so that less charges are accumulated for the same applied bias as in (b). This results in a further decrease of capacitance, due to the cross quantum capacitance discussed in the main text.